\documentclass[prl,twocolumn,superscriptaddress,showpacs,floatfix,floats]{revtex4-2}

\usepackage{epsfig}
\usepackage[centertags]{amsmath}
\usepackage{color,hyperref}
\usepackage{upgreek}
\usepackage{subfigure}

\definecolor{grey}{rgb}{.65,.65,.65}

\newcommand{\KTO}{$\rm KTaO_3$}

\begin{document}

\title{THz-induced phonomagnetism in diamagnetic quantum paraelectric KTaO$_3$}

\author{C. Kadlec}
\affiliation{Institute of Physics, Czech Academy of Sciences, Na Slovance 2, 182 00 Prague~8, Czech Republic}
\author{F. Kadlec}
\affiliation{Institute of Physics, Czech Academy of Sciences, Na Slovance 2, 182 00 Prague~8, Czech Republic}
\author{D. Rep\v{c}ek}
\affiliation{Institute of Physics, Czech Academy of Sciences, Na Slovance 2, 182 00 Prague~8, Czech Republic}
\author{P. Ku\v{z}el}
\affiliation{Institute of Physics, Czech Academy of Sciences, Na Slovance 2, 182 00 Prague~8, Czech Republic}
\author{M. Basini}
\affiliation{Department of Physics, Stockholm University, Stockholm, Sweden}
\affiliation{Department of Physics, ETH Z$\ddot{u}$rich, Auguste-Piccard-Hof 1, 8093 Z$\ddot{u}$rich, Switzerland}
\author{J.-C. Deinert}
\affiliation{Institute of Radiation Physics, Helmholtz-Zentrum Dresden - Rossendorf (HZDR), Bautzner Landstr. 400, 01328 Dresden, Germany}
\author{S. Kovalev}
\affiliation{Institute of Radiation Physics, Helmholtz-Zentrum Dresden - Rossendorf (HZDR), Bautzner Landstr. 400, 01328 Dresden, Germany}
\affiliation{Department of Physics, TU Dortmund University, 44227 Dortmund, Germany}
\author{T. Tadano}
\affiliation{National Institute for Materials Science, Tsukuba, Ibaraki 305-0047, Japan}
\author{M. Udina}
\affiliation{CESQ-ISIS (UMR 7006), Universit\'{e} de Strasbourg and CNRS, Strasbourg, 67200, France}
\affiliation{Universit\'{e} Paris Cit\'{e}, CNRS, Laboratoire Mat\'{e}riaux et Ph\'{e}nom\`{e}nes Quantiques, 75205 Paris, France}
\author{I. Ilyakov}
\affiliation{Institute of Radiation Physics, Helmholtz-Zentrum Dresden - Rossendorf (HZDR), Bautzner Landstr. 400, 01328 Dresden, Germany}
\author{T.V.A.G de Oliveira}
\affiliation{Institute of Radiation Physics, Helmholtz-Zentrum Dresden - Rossendorf (HZDR), Bautzner Landstr. 400, 01328 Dresden, Germany}
\author{A. Ponomaryov}
\affiliation{Institute of Radiation Physics, Helmholtz-Zentrum Dresden - Rossendorf (HZDR), Bautzner Landstr. 400, 01328 Dresden, Germany}
\author{A. Arshad}
\affiliation{Institute of Radiation Physics, Helmholtz-Zentrum Dresden - Rossendorf (HZDR), Bautzner Landstr. 400, 01328 Dresden, Germany}
\author{A. Maia}
\affiliation{Institute of Physics, Czech Academy of Sciences, Na Slovance 2, 182 00 Prague~8, Czech Republic}
\author{S. Bonetti}
\affiliation{Department of Physics, Stockholm University, Stockholm, Sweden}
\affiliation{Department of Molecular Sciences and Nanosystems, C\`{a} Foscari University of Venice, Venice, Italy}
\author{S. Kamba}
\thanks{Author to whom correspondence should be addressed}
\email[e-mail: ]{kamba@fzu.cz}
\affiliation{Institute of Physics, Czech Academy of Sciences, Na Slovance 2, 182 00 Prague~8, Czech Republic}

%\coloruse
\begin{abstract}
  The current efforts striving to develop new ways of data manipulation
  are aimed at ultrafast control of magnetization in magnetic materials, as well
  as at inducing magnetic moments in diamagnetics. We demonstrate that in the
  diamagnetic quantum paraelectric KTaO$_3$, the electric field of circularly
  polarized THz pulses with an amplitude of $\sim 300\,$kV/cm induces a
  transient magnetic-like response by resonantly exciting its degenerate
  soft polar phonon. This phonon-mediated response was measured using the
  THz pump---optical probe technique via the time-resolved magneto-optic Faraday
  effect. Our detection scheme was set up to cancel out the major part of the
  electro-optic Kerr effect which also usually significantly contributes to the
  transient response. The Kerr-effect-related signal was further suppressed by
  subtracting the experimental data related to oppositely circularly polarized
  THz radiation. Thus, we were able to unambiguously identify a
  temperature-dependent magnetic-like behavior of the KTaO$_3$ crystal
  manifested by the extracted Faraday rotation. We developed a theoretical
  model describing well quantitatively the measured curves of the transient
  Faraday effect signal. However, their amplitudes exhibit an unexpected
  temperature dependence, which might be a key to a deeper understanding of the
  observed phonomagnetic effect.
\end{abstract}
\date{\today}
%\pacs{75.85.+t, 75.40.Gb, 78.30.-j}

\maketitle

In the context of ongoing efforts to bring up novel efficient mechanisms of data manipulation and operation for future uses in data storage and computational applications, much attention has been paid to interactions of light with magnetic order, especially on an ultrafast scale. Indeed, it is well known that ultrashort laser pulses allow for a very fast control of magnetic states in solids. Previously, magnetic phase transitions induced by ultrafast laser heating were studied \cite{beaurepaire96}. Later, non-thermal control of magnetization by instantaneous photomagnetic pulses was used \cite{kimel05}. In this case, circularly polarized fs laser pulses excited and coherently controlled the spin dynamics in magnets via the inverse Faraday effect, generating an effective magnetic field whose direction depended on the handedness of the circularly polarized laser beam.

More recently, transient changes in crystal structure or in magnetic order were achieved by non-linear phonon coupling  and/or by phonomagnetic effects \cite{forst11}. It was demonstrated that resonant pumping of polar phonons by intense THz (or infrared) radiation may trigger a non-linear coupling with another phonon of different symmetry (e.g. a Raman-active phonon), which in turn can cause a transient change in the crystal structure or in the magnetic order \cite{forst2011driving,fechner2018magnetophononics,gu2018nonlinear,disa20,afanasiev2021ultrafast,stupakiewicz2021ultrafast,disa2023photo}. Metal-insulator phase transitions were induced in this way \cite{rini2007control}, and even a non-equilibrium superconducting phase near room temperature was observed \cite{fausti2011light,nicoletti16,mitrano2016possible,rowe2023resonant}. In 2017, Nova \textit{et al.}\cite{nova17} excited two orthogonally polarized phonons near 17 THz in antiferromagnetic ErFeO$_3$ using intense infrared radiation polarized along the [110] crystallographic direction. They have shown that non-linear mixing of two orthogonally polarized phonons enhances magnetization and excites the magnon at 0.75 THz in this material \cite{nova17}. Juraschek \textit{et al.}\cite{Juraschek17a,Juraschek19} developed their reasoning and proposed to pump \textit{degenerate} phonons in non-magnetic substances using circularly polarized THz radiation in order to induce a quasi-static magnetic moment as a consequence of a circular motion of atoms. The authors calculated the orbital magnetic moments of axial phonons in 33 different compounds and predicted the largest magnetic moments in CsH, SrTiO$_3$ and \KTO\ \cite{Juraschek19}. Based on the assumption that circularly polarized phonons with time-dependent polarization $\textbf{P}(t)$ lead to a magnetization $\textbf{M}\propto \textbf{P} \times \partial_{t}\textbf{P}$, the concept was named dynamical multiferroicity\cite{Juraschek17a,dunnett2019dynamicMFE}.

The possibility to generate magnetization via orbital magnetic moment
$\mu_{\mathrm{ph}}$ of coherently excited circularly polarized phonons
\cite{Juraschek17a,Juraschek19} was recently confirmed by Basini \textit{et al.}
\cite{basini24Nature} who drove the degenerate mode in SrTiO$_3$ near room
temperature. The peak of the pumping spectrum was set at 3 THz, coinciding with
the phonon frequency. Their measurements yielded two remarkable results:

1. The measured response was four orders of magnitude higher than the
theoretically expected value. This also suggests the phonon-induced
magnetic moment $\mu_{\mathrm{ph}}$ to be four orders of magnitude higher than predicted in Ref. \cite{Juraschek19}. To explain the discrepancy, various mechanisms of transfer of angular momentum from the phononic to the electronic degree of freedom were proposed \cite{Geilhufe22,shabala2024phonon,sellati2605light}. Merlin even proposed that the large magnetic-like signal is generated by a non-Maxwellian field that breaks the time-reversal symmetry \cite{merlin2024unraveling,merlin2025magnetophononics}. 

2. Their detected signal featured two maxima, at 0.6 and 6\,THz. The
maximum at 6\,THz (i.e., twice the pumping THz frequency) was mostly
caused by the THz-field-induced electro-optic Kerr effect (TKE) associated with
the non-linear polarization $P_{i,\mathrm{NL}} = \varepsilon_0
\chi^{(3)}_{ijkl}E_{j}T_{k} T_{l}$ (here $\chi^{(3)}_{ijkl}$ is the third-order
non-linear electric susceptibility and $E_{j}$, $T_{k,l}$ are electric field
components of the probing optical and pumping THz electromagnetic waves,
respectively) \cite{basini24Nature,Basini_PRB24}. The peak at 0.6 THz was
attributed to a combination of the TKE and the THz-field-induced magneto-optic Faraday effect (TFE).

In this article, we present results of measurements of the THz-field induced
magneto-optical TFE and electro-optical TKE on a single crystal of \KTO. \KTO\
has large Born effective charges and it exhibits a temperature
dependence of the soft mode (SM) phonon frequency very similar to that of
SrTiO$_3$ \cite{Suppl,Vogt95}. The advantage of \KTO\ over SrTiO$_3$ is
that its SM damping is very low ($\leq$~60~GHz below 50\,K) and its crystal structure remains cubic down to absolute zero. This
enables measurements at low temperatures where the SM potential is shallow so an
efficient coupling with the THz field is expected, leading to higher phonon
amplitudes and higher values of $\mu_{\mathrm{ph}}$. Also, according to
theoretical analysis \cite{Juraschek19}, the magnetic moments owing to the SMs in
KTaO$_3$ and SrTiO$_3$ should be the highest among all perovskites.
Being aware of the difficulties caused by the superposition of the TFE with the
concomitant TKE, we optimized the experimental setup to reject this unwanted
contribution without any need of material's response modelling. Thus we
were able to identify unambiguously the magnetic-like TFE response directly from the acquired data.

A (001)-cut \KTO{} single crystal in the form of an optically polished
parallelepiped with dimensions approximately $7\times8\times0.63\,\rm mm^3$ and
edges oriented along [100] and [010] axes was used (these directions also
coincided with $x$ and $y$ axes of our laboratory frame). Its quantum-paraelectric behavior was verified through the temperature dependence of the dielectric permittivity, see Fig.\ S2 in Supplementary information (SI) \cite{Suppl}. 

The pump-probe experiments were performed on the TELBE high-field THz facility
at the Helmholtz-Zentrum Dresden-Rossendorf \cite{green2016, helm2023elbe} where
a linear electron accelerator generates tunable intense multi-cycle THz pulses
at a repetition rate of 50\,kHz. We used pulses with an energy of
$\sim$10\,$\mu$J, an electric field amplitude of $\sim$300 kV/cm in the focus, a
central frequency of $\Omega_0=0.7\,\mathrm{THz}$ and a
full-width-at-half-maximum (FWHM) of $\sim$65\,GHz pertaining to the power spectrum (see Fig.
S5 in \cite{Suppl}). Experiments were performed with an either linearly
or circularly polarized THz pump beam. In the former case, the pump
polarization was oriented along the [100] axis of \KTO; only TKE can be induced.
In the latter case, a quartz quarter-wave plate was used to transform the
linearly polarized pumping pulse into a circular one; in fact, it involved an
elliptical imperfection, namely a difference of about 20\,\% in
perpendicular electric-field amplitudes along $x$ and $y$ axes. We carried out
experiments with both right-hand-polarized (RHP) and left-hand-polarized (LHP)
pump. For probing we used time-synchronized femtosecond Ti:sapphire laser pulses
centered at the wavelength of $\omega_{\rm p}=800$\,nm and linearly
polarized along the [110] crystallographic direction. Polarization-sensitive
balanced photodetection was employed to detect the THz-pump-induced optical
component $F$ polarized along the $[1\bar{1}0]$ direction,
perpendicularly to the incident probe field $E$, as described in End Matter (Fig.
\ref{fig:Detection}). The data traces $F(t)$ were recorded as a function
of the pump-probe delay $t$; the photodiodes had a broadband
sensitivity, so they measured both the transient signal induced at the
incident probe beam frequency $\omega_p$ (main band) centered in
the THz spectrum at
$\Omega\approx 0$ and the one induced at the sidebands, $\omega_p\pm
2\Omega_0$, which appeared in the traces as an oscillation at $\Omega\approx 2\Omega_0$. The experiments were performed in transmission geometry at temperatures down to 11\,K. A complete experimental schema with more technical details is provided in SI (Fig. S4 in \cite{Suppl}).

It was shown experimentally and theoretically \cite{basini24Nature,Basini_PRB24} that in THz pump---optical probe experiments, the TKE always generates an electromagnetic contribution, for both linearly and circularly polarized THz pump. Therefore, for studying efficiently the TFE, it is desirable to conceive an experimental scheme minimizing the TKE contribution. Compared with the experimental setup of Ref.~\cite{basini24Nature}, in our setup, we turned the incident polarization of the probe beam by 45$^\circ$ to the [110] direction which allowed us to avoid a half-wave plate between the sample and the photodiodes. Indeed, this scheme allowed us to efficiently cancel out the TKE contribution for the circularly polarized pump, as briefly described in End Matter and in detail in SI \cite{Suppl} by solving the nonlinear wave equation. In fact, the phase of the TKE-induced contribution is essentially $\pi/2$ shifted compared to the phase of the unperturbed probe wave, making the detection insensitive to the first-order TKE contribution. Even more importantly, for this particular experimental arrangement, the TKE contribution to the main band vanishes. Finally, the differential signal from balanced photodiodes can be written  as
\begin{equation}
  \Delta\Gamma_\mathrm{RHP/LHP}=
  \pm\Delta\Gamma_{\mathrm{TFE}}+\delta\Gamma_{\mathrm{TKE}}\,,
\end{equation}
where the right- and left-hand-polarized pump is denoted by RHP and LHP, respectively, and
they induce a positive and negative TFE signal $\Delta\Gamma_{\mathrm{TFE}}$. In contrast, the
sign of the second-order residual TKE contribution
$\delta\Gamma_{\mathrm{TKE}}$ is independent of the pump handedness (see End Matter). After the subtraction $\Delta\Gamma_\mathrm{dif}=\Delta\Gamma_\mathrm{RHP}-\Delta\Gamma_\mathrm{LHP}= 2\Delta\Gamma_{\mathrm{TFE}}$, the signal is free of the TKE, provided that the RHP and LHP pumps are indeed circularly polarized or slightly elliptically polarized with the principal axes along [100] and [010] crystallographic directions of \KTO\ (which is true in our case, as shown in Sec. IV of SI).

\begin{figure}
    \centering
    \includegraphics[width=75mm]{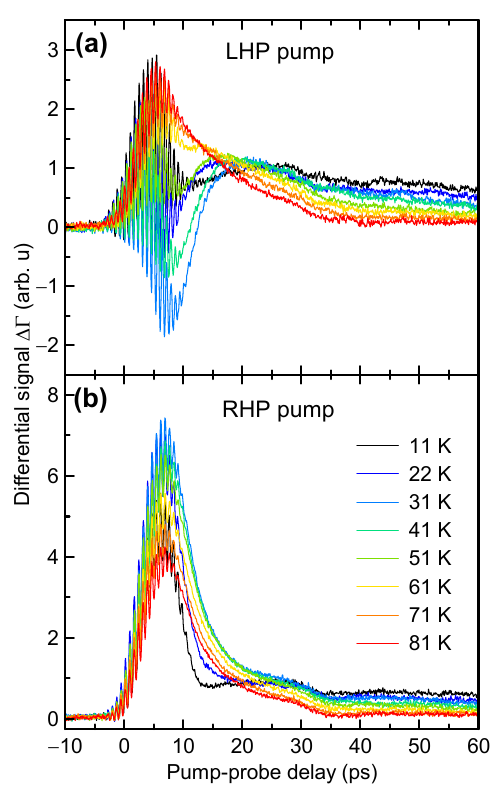}
    \caption{Time dependences of the differential signal $\Delta \Gamma$ at various temperatures in \KTO{} upon pumping with (a) LHP and (b) RHP THz pulse. The scale in arbitrary units is the same for the two plots. }
    \label{fig:time_traces}
\end{figure}

Fig. \ref{fig:time_traces} displays experimental results obtained for
nominally circularly polarized pump pulses as a function of temperature.
In the plots all curves clearly contain a quasi-static component  with
$\Omega\approx 0$ and an oscillating one at $\Omega\approx 2\Omega_0$.
Since the $2\Omega_0$ component is forbidden for TFE, it must be due to the
residual TKE, and this part of the signal provides us with an order-of-magnitude
estimate of the amplitude of the detected TKE. The quasi-static part of the
signal has a fast Gaussian-like shape between roughly 0 and 10--15\,ps and a
long-lived contribution, extending up to 30 ps for higher temperatures (81
K) and exceeding 60 ps for the lowest temperatures. The long-lived component
clearly exhibits the same sign for RHP and LHP pump at all temperatures and thus
it is supposedly the TKE residual which should vanish by subtracting
$\Delta\Gamma_\mathrm{RHP}-\Delta\Gamma_\mathrm{LHP}$. The fast Gaussian-like
component has again the same sign for the two pump polarizations at high
temperatures (61--81 K) but, upon further cooling, the sign of LHP signal
changes. The sign reversal upon the LHP to RHP switching corresponds to the
time-reversal induced by the THz pump to the crystal and thus we observe here an
onset of the TFE. Indeed, between 10 and 50 K the SM resonance frequency
increases from 0.6 to 1\,THz (see Fig. S1 in \cite{Suppl}), so
its component with a non-zero orbital moment can be resonantly driven by
the circularly polarized pump pulses inducing a net magnetization, as predicted
in Ref.~\cite{geilhufe21} and as expressed by Eqs. (S23) and (S24) in \cite{Suppl}.

\begin{figure}
\centering \includegraphics[width=75mm]{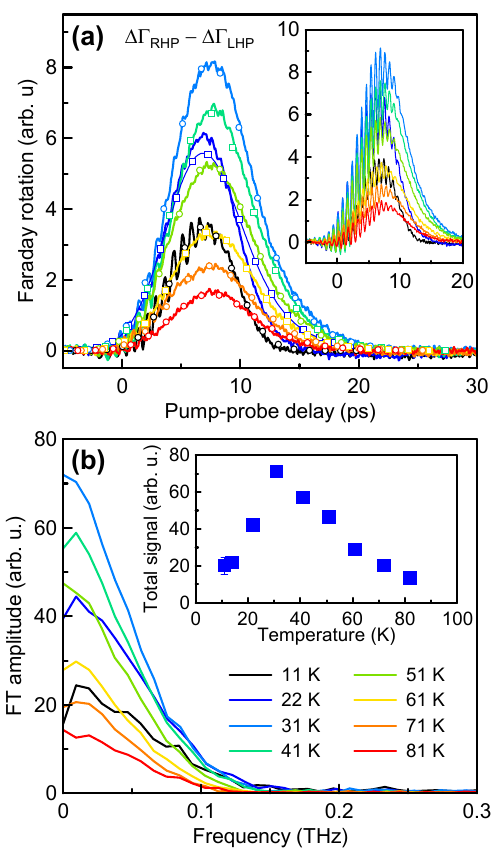}
\caption{(a) Lines: THz-induced Faraday rotation calculated as a
    difference of the signals $\Delta \Gamma_\mathrm{RHP}-\Delta
    \Gamma_\mathrm{LHP}$ as shown Fig.~\ref{fig:time_traces} at various
    temperatures. A time-shift correction was applied to the raw time-domain
    data (as described in the text). Symbols: fits of the data by the
    theoretical model given by Eq. (S47) of SI (see text for details).
    Inset: difference traces of the raw time-domain data (i.e., without the
    time-shift correction). (b) Fourier transform of the
    time-domain traces in the main panel of (a). Inset: temperature dependence
    of the integral Faraday signal.}
	  \label{fig:Faraday-new}
\end{figure}

As discussed above and in End Matter, calculating the difference between the LHP
and RHP data, the TKE contribution is subtracted and a pure TFE response
is obtained. However, the oscillations at $2\Omega_0$ have a period of
$\sim 0.7$ ps and in the TELBE setup a relative time delay (of the order of a
few hundreds of fs) between the arrivals of LHP and RHP pump pulses can arise
leading to a slightly different time origin of the parts (a) and (b) in
Fig. \ref{fig:time_traces}. This shift of the time origin may be due to
several factors, namely the rotation of the quarter-wave plate and imperfections
in parallelism of its surfaces, as well as software corrections of the
pulse time jitter in the TELBE beamline. In this context we introduced an additional
parameter, representing the time delay between the LHP and RHP data at a given
temperature. Its resulting values, not exceeding 300\,fs, were chosen so as to minimize the
$2\Omega_0$ component of the $\Delta\Gamma_{\rm dif}$ curves. This leads to time traces
which can be interpreted as due to the sole TFE effect; these are
shown in Fig. \ref{fig:Faraday-new}(a) as a function of temperature. (For
illustration, the inset shows results of subtracting 
 the raw data without any time shift.) We also stress that the time shift values
 we used do not influence the quasi-static peak originating in the TFE (see Fig.\ S7 in SI). The corresponding
 TFE spectra calculated by Fourier transformation are shown in Fig.
 \ref{fig:Faraday-new}(b). Its inset then displays the temperature dependence of
 the amplitude of the quasi-static Faraday signal. This quantity
 was calculated as an average of the two lowest-frequency values of
 the main panel at each temperature. We note that the error bar, defined as the
 difference of these values, is similar to (or smaller than) the size of the data symbols. 

\begin{figure}
	  \centering
	  \includegraphics[width=75mm]{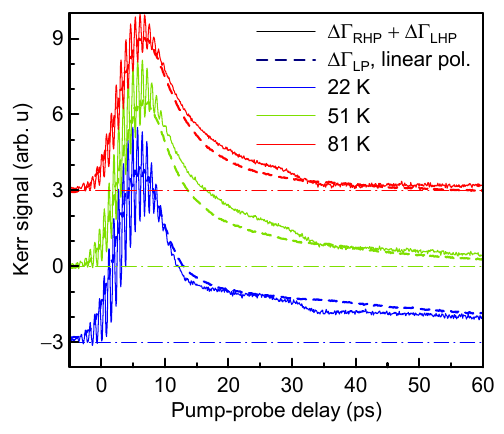}
\caption{Comparison of the sums of traces $\Delta \Gamma_\mathrm{RHP}+\Delta
\Gamma_\mathrm{LHP}$ obtained for RHP and LHP THz pumps (solid line) with
independently measured signals $\Delta \Gamma_\mathrm{LP}$ for a linearly
polarized THz pump (dashed line). The plots are vertically shifted for clarity
(thin dash-dotted lines indicate zero signal level for each part).}
	  \label{fig:Kerr-new}
\end{figure}

For a more complete analysis, we also measured the TKE
$\Delta\Gamma_\mathrm{LP}$ using a linearly polarized
THz pump beam; in this case, no magnetic contribution
can be induced. At the same time, the magnetic contribution in the TFE
measurements with a circularly polarized pump should be canceled
upon summing the time traces measured with the two circular polarizations of the
pump: $\Delta\Gamma _\mathrm{RHP}+\Delta\Gamma_\mathrm{LHP}$. A comparison
of these two types of curves for selected temperatures is shown in Fig.
\ref{fig:Kerr-new}. The long-lived component
(previously discussed in the literature in terms of THz-induced electric
polarization and correlations between polar nanoregions which
develop especially at low temperatures \cite{Li23,Cheng23}) exhibits the
same decay for both data sets. The component calculated from experiments with
circularly polarized pump shows a significantly stronger sideband
($2\Omega_0$) component since, in principle, different terms contribute to the signal in this case.

Note that the Faraday rotation $\Delta\Gamma_{\rm TFE}$ shows a non-monotonic temperature
dependence, as inferred from Fig. \ref{fig:Faraday-new}(a,b). Its values
are relatively low at 81 K and they increase with cooling down to 31\,K,
where the TFE intensity exhibits a maximum. Further cooling below 31 K
is accompanied by a decrease in the TFE signal. This is in qualitative agreement
with the theory of dynamical multiferroicity with a temperature-dependent
soft optical phonon \cite{Juraschek17a}. We performed a fit of the time-domain
Faraday traces shown in Fig.\ \ref{fig:Faraday-new}(a) using the formula (S47)
derived in SI \cite{Suppl}. For the calculations, we used a circularly polarized
time domain waveform with a Gaussian envelope, which matches very well the
experimentally measured pump waveforms (see Fig. S5). We also used the
experimentally determined temperature- and frequency-dependent linear dielectric
susceptibility $\chi(\Omega;T)$ and the related THz refractive index
$N(\Omega;T)$ of \KTO\ (Fig.\ S3 in \cite{Suppl}); these quantities enter Eq.\
(S47) through the functions $V$ and $W$ defined by Eqs. (S32) and (S33). The
only fitting parameters were the experimental time origin and the amplitude of
the trace for each temperature, $D(T)$, introduced in Eq. (S47). The fit yielded an excellent agreement between the experimental time traces and the
theoretical model, except for the measurement at 22 K 
[see Fig.\ \ref{fig:Faraday-new}(a)]. We attribute this small discrepancy to the fact
that at this temperature the SM frequency matches the bandwidth of the pump
pulse and, due to these resonance conditions, some additional higher-order
nonlinear effects can take place. The temperature dependence of the amplitude
of the traces $D(T)$ follows an unexpected trend which is discussed in detail in End Matter.

To conclude, we clearly demonstrated a phonomagnetic response of \KTO\ by means
of experiments where the soft phonon was coherently driven by strong-field
circularly polarized THz pulses and a magnetic-like response was detected via the time-resolved magneto-optical Faraday effect.
 A proper experimental setup and an in-depth theoretical analysis of nonlinear
 interactions enabled us to eliminate the influence of the electro-optic Kerr
 effect, which was predominant in a similar earlier experiment on
 SrTiO$_3$ \cite{basini24Nature}. Our approach reveals a straightforward method of identifying the Faraday phonomagnetic effect, which will be very useful for
 further research in chiral phonons and phonomagnetism. Although our model is very appropriate for quantitatively
 describing the mechanism of inducing the magnetic-like response of \KTO\ and
 the shapes of the concomitant Faraday effect transiently induced by strong THz
 pulses, it does not explain the observed temperature dependence of the amplitude of the Faraday effect. This provides an opportunity for follow-up studies
 which should bring a deeper understanding of the clearly demonstrated
 phonomagnetic effect in diamagnetic quantum paraelectric KTaO$_3$.

\begin{acknowledgments}
  \textit{Acknowledgments---}This work was supported by the Czech Science Foundation (Project No. 24-10791S) and by project TERAFIT - CZ.02.01.01/00/22\_008/0004594 co-financed by European Union and the Czech Ministry of Education, Youth and Sports. Parts of this research were carried out at ELBE at the Helmholtz-Zentrum Dresden - Rossendorf e. V., a member of the Helmholtz Association. MB and SB acknowledge support from the Knut and Alice Wallenberg Foundation (Grant No. 2019.0068) and from the SNSF Ambizione project PZ00P2\_216089.
\end{acknowledgments}

%\section*{Data availability}

\textit{Data availability---}Data supporting the findings of this article are publicly available \cite{Zenodo}.

%\bibliographystyle{apsrev4-2}
%%\bibliographystyle{unsrt}
%\bibliography{KTO.bib}

\begin{thebibliography}{35}
\bibitem [{Zen()}]{Zenodo}%
  \BibitemOpen
  \href@noop {} {}\bibinfo {howpublished} {Dataset on Zenodo},\ \bibinfo {note}
  {link to be provided}\BibitemShut {NoStop}%
\end{thebibliography}

\onecolumngrid
\section*{End Matter}
\twocolumngrid

\subsection{Principle of detection}
\begin{figure}
	  \centering
	  \includegraphics[width=\columnwidth]{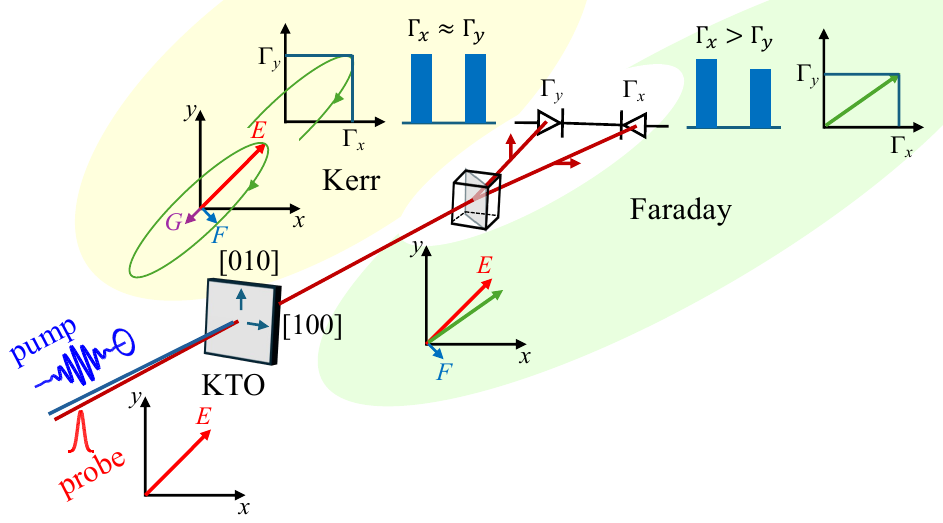}
\caption{Detection scheme. Incident probe beam:
$\mathbf{E}=(E,E,0)/\sqrt{2}$; output probe beam:
$\mathbf{E}_{\mathrm{probe}}=(E-G+F,E-G-F,0)/\sqrt{2}$. Vectors $\mathbf{E}$,
$\mathbf{F}$ and $\mathbf{G}$ indicate the polarization state of the probe beam
after the interaction with the sample. In the case of the Kerr effect the $F$
and $G$ are phase shifted by $\pi/2$,  the output polarization is elliptical and
the measured difference signal is proportional to $\Re (G^\ast F)$, vanishing in
the first order. In the case of the Faraday effect, $G=0$ and $F$ is in phase
with $E$, thus the difference signal is $\Re (E^\ast F)$.  }
	  \label{fig:Detection}
  \end{figure}

In Fig. \ref{fig:Detection} we show the principle of detection. In the
laboratory frame $xyz$ (which coincides here with the crystallographic axis
frame of \KTO), the incident optical probing pulse $\mathbf{E}$ is linearly
polarized along the [110] direction, i.e., $\mathbf{E}=(E,E,0)/\sqrt{2}$. The THz-pump-induced optical field may have components polarized both along [110] (denoted as $G$ in Fig. \ref{fig:Detection}) and along $[1\bar{1}0]$ (denoted as $F$), i.e., the total field transmitted through the sample reads $\mathbf{E}_{\mathrm{probe}}=(E-G+F,E-G-F,0)/\sqrt{2}$, $F,G\ll E$. Moreover, owing to the frequency mixing in the nonlinear interaction, each THz-pump induced polarization component may consist of a wave oscillating at the original probe frequency $\omega_\mathrm{p}$ (main band) and of a wave oscillating at shifted frequencies $\omega_\mathrm{p}\pm 2\Omega_0$ (sidebands), where $\Omega_0\approx 0.7\, \mathrm{THz}$ is the central frequency of the THz pump pulse. These waves will be denoted as $F_0$, $G_0$ (main band) and $F_2$, $G_2$ (sidebands). The detection part consists of a Wollaston polarizer, which separates the probe beam into two parts, one with horizontal ($E_{\mathrm{probe},x}$) and the other with vertical ($E_{\mathrm{probe},y}$) polarization, and of a pair of balanced photodiodes, which deliver the difference signal $\Delta\Gamma$ between the two beams:
\begin{multline}
    \Delta\Gamma = \Gamma_x-\Gamma_y \propto\left |E_{\mathrm{probe},x}\right
    |^{2}-\left |E_{\mathrm{probe},y}\right |^{2}=\\ 2\Re (E^\ast F) - 2\Re (G^\ast F)\,.
    \label{eq_DGamma}
\end{multline}
Note that $\Re(A^\ast B)=0$ if $A$ and $B$ are mutually $\pi/2$ shifted. TKE and
TFE have different origins and therefore the polarization state of the output
probe beam is different in the two cases. In SI \cite{Suppl} we solve the
appropriate wave equation for both linearly and circularly polarized THz pump
pulses. In principle, TKE leads to an elliptical polarization of the
output probe beam, which means that the component $F$ is phase shifted with
respect to $E$. We further show that in the case of a circularly polarized
pump (a) $F_0=0$ [see Eq.\ (S14)], (b) the phase shift of $F_2$ and $G_0$ is
practically equal to $\pi/2$, and (c) the phase shift is zero for $G_2$ (see
also Table \ref{tab:F_G_components} for a summary of these properties). In this
case, the first order signal $\Re (E^\ast F)$ vanishes and only the second order
contribution, $\Re (G_0^\ast F_2)$, can be detected (this signal is caused by a
very small tilt of the principal axes of the polarization ellipse with respect
to $[110]$ direction owing to the $\pi/2$-shifted $G$-component). Note that this
term contributes merely to the sidebands. The main band component induced by the
Kerr effect might appear in the measured signal only due to imperfect
polarization states of the pump and probe beams, due to the finite bandwidth of
the probe pulse (i.e., finite temporal pulse length), or due to a very
weak non-phase-matched term (see SI for further details).
\begin{table}
\begin{tabular}{|c|c|c|c|} \hline
   \textbf{Component} & \textbf{Main band} & \textbf{Sidebands} &  \textbf{RHP vs. LHP} \\
  \hline
  \multicolumn {4}{|l|} {\quad TKE -- linear pump} \\
\hline
    $F$ & $i F_0\neq 0$ & $i F_2\neq 0$  & NA \\
\hline
    $G$ & $i G_0\neq 0$ & $i G_2\neq 0$  & NA \\
\hline
  \multicolumn {4}{|l|} {\quad TKE -- circular pump} \\
\hline
    $F$ & $ F_0= 0$ & $i F_2\neq 0$ & $F_{2,\mathrm{RHP}}= F_{2,\mathrm{LHP}}$\\
\hline
    $G$ & $ i G_0\neq 0$ & $G_2\neq 0$ & $G_{0,\mathrm{RHP}}= G_{0,\mathrm{LHP}}$\\
  \hline
  \multicolumn {4}{|l|} {\quad TFE -- circular pump} \\
\hline
    $F$ & $ F_0\neq 0$ & $F_2 = 0$ & $F_{0,\mathrm{RHP}}= -F_{0,\mathrm{LHP}}$\\
\hline
    $G$ & $ G_0= 0$ &  $G_2= 0$ &  NA\\
\hline
\end{tabular}
 \caption{Properties of THz-pump-induced components (see Fig. \ref{fig:Detection}) for various interactions. The imaginary unit denotes an approximate $\pi/2$ phase shift with respect to the incident probe field; its absence represents zero phase shift. All the theoretical details are explained in SI \cite{Suppl}.}\label{tab:F_G_components}
\end{table}
In the case of TFE, $F_0$ oscillates in phase with $E$ and $F_2=0$, $G=0$ (see
SI for the derivation); the measured signal  consists of the main band only, and
it is proportional to $\Re (E^\ast F_0)$, see also Table \ref{tab:F_G_components}. In this way our experimental scheme leads to a linear Faraday signal and it rejects the Kerr component as much as possible.

Next we calculate the difference between the signals for RHP and LHP pump. The
rejection of the Kerr contributions is then inherent due to the symmetry of the
RHP and LHP induced signals as shown in the last column of Table
\ref{tab:F_G_components}. It follows that for TKE
$\Delta\Gamma_\mathrm{RHP}-\Delta\Gamma_\mathrm{LHP}= 0$, while for TFE
$\Delta\Gamma_\mathrm{RHP}-\Delta\Gamma_\mathrm{LHP}= 2
\Delta\Gamma_\mathrm{TFE}$.

Note that if the probe beam impinges on the sample linearly polarized
along a crystallographic axis of \KTO\, e.g. along [010], and the half-wave
plate is positioned only after the sample, which was the case in
\cite{basini24Nature}, the main difference compared to our current case
shown in Fig.\ \ref{fig:Detection} lies in the properties of TKE with a circular
pump. Indeed, as shown in SI, in this case, $F_0\neq 0$, see Eq.\ (S12) in
\cite{Suppl}, which leads to a more intense residual pollution of the Faraday
signal by the main band of TKE. Moreover, another consequence of this
experimental setup is that for the circular pump, $F_2$ and $G_2$ will feature a
general phase shift with respect to the unperturbed probe wave; this phase shift
will depend on the magnitudes of various nonlinear susceptibility coefficients. 

\subsection{Origin of the $D(T)$ dependence}

\begin{figure}
	  \centering
	  \includegraphics[width=75mm]{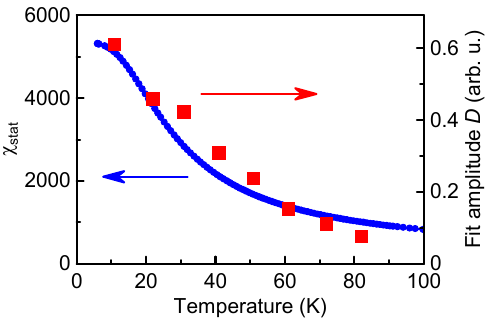}
\caption{Comparison of temperature dependencies of the linear dielectric susceptibility value of our \KTO\ crystal measured at 1\,MHz (blue line, left scale) and of the amplitude $D$ of the fit of the Faraday rotation.}
	  \label{fig:chim}
\end{figure}

The temperature dependence of the amplitude
of the traces $D(T)$ is shown in Fig.\ \ref{fig:chim}. We emphasize that the
temperature dependence of the Faraday effect due to the temperature
evolution of the \KTO\ susceptibility is
included in our model [see Eq. (S47) in Ref.~\onlinecite{Suppl}], and if the susceptibility were the only
source of the temperature variation of the Faraday signal, $D(T)$ would be
constant. Clearly, the observed shape of $D(T)$ exceeds the scope of our model. Roughly speaking, as the temperature drops from 80 to
11 K, the amplitude of the observed signal becomes eight times stronger than
predicted, e.g., by the simplified expression in Eq. (S48). Interestingly, this
dependence is similar to the temperature variation of the static linear electric
susceptibility in \KTO\, which also increases monotonically towards
0\,K (see Fig.\ \ref{fig:chim}). 
Below we discuss the possible origins of the observed shape of the $D(T)$
dependence (see Fig.~\ref{fig:chim}).

(i) It is unlikely that it comes from the temperature dependence of the
parameter $g$ in (S47) which is proportional to the Verdet constant.
As a rule, in diamagnets, the temperature dependence of the Verdet constant is very weak, as it is mostly due to the temperature dependence of the optical refractive index \cite{Williams91}. 

(ii) We theoretically estimated the temperature dependence of the gyromagnetic
ratio (parameter $\gamma$), see the description in SI \cite{Suppl} and Fig. S10 therein. We found
that $\gamma$ can vary at most by 5\% in the temperature range between 0 and 100
K, far less than in the observed temperature dependence of the
amplitude
$D(T)$.

(iii) It has been suggested in recent papers \cite{geilhufe21,Geilhufe22} that
an orders-of-magnitude enhancement of the phonomagnetic effect could be expected
if an additional related contribution by free electrons is taken into account.
However, it appears that, in order to induce an appreciable effect, the
concentration of free electrons should be sufficiently high to shift the Fermi
level very close to (or even above) the conduction band edge (see Fig.\ 4 in
\cite{Geilhufe22}). In principle, conduction band electrons may be present in
the system due to ionized oxygen vacancies or due to multiphoton ionization by
strong-field THz pulses. In our case, the steady-state conductivity of our
insulating \KTO\ crystal is negligible and so must be the concentration of
electrons. Dynamical generation of conduction band electrons by strong  pulses
at 0.7\, THz would require at least 30-photon absorption to overcome the energy difference of 0.1--0.2 eV for oxygen vacancies levels. So this mechanism seems also very unlikely.

(iv) One can consider a contribution of the nonlinear polarization component in
the formula (S21). In the case we take into account the third-order (Kerr-like)
polarization $\mathbf{P}_\mathrm{NL}^{(3)}$, leading to an additional
phonomagnetic contribution, such as $\mathbf{B}^{(3)}=\gamma \,
\dot{\mathbf{P}}\times \mathbf{P}_\mathrm{NL}^{(3)}$, and we carry out the
calculation of the Faraday signal, we find again a vanishing $2\Omega_0$
component and a non-zero $\Omega\approx0$ component featuring an additional
susceptibility term, see Eq. (S50), as required by the temperature dependence shown
in Fig.\ \ref{fig:chim}. However, since this nonlinear contribution to the
phonomagnetic field is additive with the linear one, it would have to dominate in order to imprint the required temperature dependence to the measured signal (cf.\ Fig.\ S6). At the same time, the third order polarization would also lead to a $4\Omega_0$ component, which was not observed.

(v) Recent papers by Merlin
\cite{merlin2024unraveling,merlin2025magnetophononics} suggest that, by
symmetry, phenomena analogous to the phonomagnetic effect discussed here may arise from a term $T\times T^\ast$, which has apparently the same symmetry as the polarization-induced magnetic field given by (S21), and which is non-zero for the circular THz polarization. However, the magnitude and temperature behavior of the related coupling constant is completely unknown.   

\end{document}